\documentstyle[preprint,12pt,aps]{revtex}

\begin{document}

\title{Exact solution of the two-dimensional Dirac oscillator}

\author{V\'{\i}ctor M. Villalba}
\address{Centro de F\'{\i}sica\\
Instituto Venezolano de Investigaciones Cient\'{\i}ficas, IVIC\\
Apdo 21827, Caracas 1020-A, Venezuela}

\maketitle
\begin{abstract}
In the present article we have found the complete energy spectrum and
the
corresponding eigenfunctions of the Dirac oscillator in two spatial
dimensions.
We show that the energy spectrum depends on the spin of the Dirac
particle.
\end{abstract}

\pacs{03.65,-w, 11.10Qr}

Recently, Moshinsky and Szczepaniak \cite{Moshinsky} have proposed a new
type of interaction in the Dirac equation which, besides the momentum,
is
also linear in the coordinates. They called the resulting the Dirac
equation
the Dirac oscillator because in the non-relativistic limit it becomes a
harmonic oscillator with a very strong spin-orbit coupling term. Namely,
the
correction to the free Dirac equation
\begin{equation}
\label{one}i\frac{\partial \Psi _c}{\partial t}=(\beta {\bf \gamma
p}+\beta
m)\Psi _c
\end{equation}
reads
\begin{equation}
\label{two}{\bf p}\rightarrow {\bf p}-im\omega \beta {\bf r}
\end{equation}
after substituting (\ref{two}) into (\ref{one}) we get an Hermitian
operator
linear in both {\bf p} and {\bf r. }Recently, the Dirac oscillator has
been
studied in spherical coordinates and its energy spectrum and the
corresponding eigenfunctions have been obtained \cite{Benitez}. A
generalization of the one-dimensional version of the Dirac oscillator
has
been proposed by Dom\'{\i}nguez-Adame. In this case, the modification of
the
free
Dirac equation, written in cartesian coordinates, is made by means of
the
substitution $m\rightarrow m-i\gamma ^0\gamma ^1V(x_1).$ Obviously, for
$%
V(x_1)=m\omega x_1$ we have the standard Dirac oscillator. Here, as well
as
for the three dimensional Dirac oscillator\cite{Benitez}, bound states
are
present.

An interesting framework for discussing the Dirac oscillator is a 2+1
space-time. The absence of a third spatial coordinate permits a series
of
interesting physical and mathematical phenomena like fractional
statistics%
\cite{Wilczek} and Chern-Simmons gauge fields among others. Since we are
interested in studying the Dirac oscillator in a two-dimensional space,
a
suitable system of coordinates for writing the harmonic interaction are
the
polar $\rho $ and $\vartheta $ coordinates. In this case the radial
component of the modified linear momentum takes the form: $p_\rho
-im\omega
\beta \rho .$ It is the purpose of the present paper to analyze the
solutions and the energy spectrum of the 2+1 Dirac oscillator expressed
in
polar coordinates.

One begins by writing the Dirac equation (\ref{one}) in a given
representation of the gamma matrices. Since we are dealing with two
component spinors it is convenient to introduce the following
representation
in terms of the Pauli matrices\cite{Hagen}

\begin{equation}
\label{rep}\beta \gamma _1=\sigma _1,\ \beta \gamma _2=s\sigma _2,\
\beta
=\sigma _3
\end{equation}
where the parameter $s$ takes the values $\pm $$1$ ($+1$ for spin up and
$-1$
for spin down). Then, the Dirac (\ref{uno}) equation written in polar
coordinates reads

\begin{equation}
\label{three}iE\Psi =\left[ \sigma ^1\partial _\rho +\sigma ^2(\frac{%
ik_\vartheta s}\rho -m\omega \rho )+i\sigma ^3m\right] \Psi
\end{equation}
with,%
$$
\Psi =\Psi _0(\rho )e^{i(k_\vartheta \vartheta -Et)}
$$
where the spinor $\Psi $ is expressed in the (rotating) diagonal gauge,
related to the cartesian (fixed) one by means of the transformation
$S(\rho
,\vartheta )$\cite{Shishkin}
\begin{equation}
\Psi _c=S(\rho ,\vartheta )^{-1}\Psi
\end{equation}
where the matrix transformation $S(\rho ,\vartheta )$ can be written as
follows
\begin{equation}
S(\rho ,\vartheta )=\frac 1{\sqrt{\rho }}\exp (-i\frac \vartheta 2\sigma
^3)
\end{equation}
noticing that $S(\rho ,\vartheta )$ satisfies the relation
\begin{equation}
S(\rho ,\vartheta +2\pi )=-S(\rho ,\vartheta )
\end{equation}
we obtain,
\begin{equation}
\Psi (\vartheta +2\pi )=-\Psi (\vartheta )
\end{equation}
so we have $k_\vartheta =N+1/2$, where $N$ is an integer number

Using the representation (\ref{rep}), the spinor equation (\ref{three})
can
be written as system of two first order coupled differential equations,
\begin{equation}
\label{uno}i(E-m)\Psi _1=(\frac d{d\rho }+\frac{k_\vartheta s}\rho
-im\omega
\rho )\Psi _2
\end{equation}
\begin{equation}
\label{dos}i(E+m)\Psi _2=(\frac d{d\rho }-\frac{k_\vartheta s}\rho
+im\omega
\rho )\Psi _1
\end{equation}
where
\begin{equation}
\Psi _0=\pmatrix{\Psi _1 \cr
\Psi _2\cr
}
\end{equation}
substituting (\ref{dos}) into (\ref{uno}) and vice-versa we arrive at

\begin{equation}
\label{sist}\left[ \frac{d^2}{d\rho ^2}-\frac{(k_\vartheta
s)(k_\vartheta
s\mp 1)}{\rho ^2}+m\omega (2k_\vartheta s\pm 1)-m^2\omega ^2\rho
^2+(E^2-m^2)\right] \pmatrix{\Psi _1 \cr
\Psi _2\cr
}=0
\end{equation}
It is not difficult to see that the solution of of the second order
equation
(\ref{sist}) for $\Psi _1$ can be expressed in terms of associated
Laguerre
polynomials $L_k^s(x)$\cite{Kamke,Abramowitz} as follows
\begin{equation}
\label{Psi}\Psi _1=c_1\exp (-x/2)x^{\frac 12(1/2+\mu )}L_n^\mu (x)
\end{equation}
where we have made the change of variables
\begin{equation}
x=m\omega \rho ^2
\end{equation}
where $\mu $ satisfies the relation
\begin{equation}
\mu =\pm (k_\vartheta s-1/2)
\end{equation}
and the natural number $n$ satisfies the relation
\begin{equation}
\label{ene}\frac{E^2-m^2}{m\omega }+(1\mp 1)(2k_\vartheta s-1)=4n
\end{equation}
Since the function $\Psi _1$ must be regular at the origin, we obtain
that
the sign of $\mu $ in (\ref{Psi}) is determined by the sign of $s.$ In
fact,
for $k_\vartheta s>0$ we have that $\Psi _1$ reads
\begin{equation}
\label{sol}\Psi _1=c_1\exp (-x/2)x^{k_\vartheta s/2}L_n^{k_\vartheta
s-1/2}(x)
\end{equation}
substituting (\ref{sol}) into (\ref{dos}) we arrive at
\begin{equation}
\label{sol2}\Psi _2=2ic_1\frac{(m\omega )^{1/2}}{E+m}\exp
(-x/2)x^{(k_\vartheta s+1)/2}L_{n-1}^{k_\vartheta s+1/2}(x)
\end{equation}
where $c_1$ is an arbitrary constant.

\noindent Analogously, we obtain that the regular solutions for
$k_\vartheta
s<0$ are,
\begin{equation}
\Psi _2=c_2\exp (-x/2)x^{-k_\vartheta s/2}L_n^{-k_\vartheta s-1/2}(x)
\end{equation}
\begin{equation}
\Psi _1=2ic_2\frac{(m\omega )^{1/2}}{E+m}\exp (-x/2)x^{(1-k_\vartheta
s)/2}L_n^{1/2-k_\vartheta s}(x)
\end{equation}
where $c_1$ is  a normalization constant The expression (\ref{ene}) can
be
rewritten as follows
\begin{equation}
\label{spec}E^2-m^2=4\left[ n-\Theta (-k_\vartheta s)(k_\vartheta
s-1/2)\right] m\omega
\end{equation}
where $\Theta (x)$ is the Heaviside step function. Then from the
relation (%
\ref{spec}) it is clear that the energy spectrum of the 2+1 Dirac
oscillator
depends on the value of s. Notice that for positive values of
$k_\vartheta s$
there is not degeneration of the energy spectrum. For $k_\vartheta s<0$
we
observe that all the states with $(n\pm l,$ $k_\vartheta s-1/2\pm l),$
where
$l$ is an integer number, have the same energy. In this direction there
are
some differences with the spherical Dirac oscillator \cite{Benitez}.
Despite
in both cases bound states are obtained, for the 2+1 Dirac oscillator
the energy
spectrum is degenerate only for negative values of $k_\vartheta s.$ In
order
to get a deeper understanding of the dependence of the energy spectrum
on the
spin we can take the nonrelativistic limit of the Dirac equation
(\ref{three}).
In order to do that, it is advisable to work with Eq. (\ref{sist}). The
Galilean limit is obtained by setting $E=m+\epsilon$, and considering
$\epsilon<<m$. Taking into account that the first two terms in
Eq.(\ref{sist})
are associated with the operator $\-P^{2}$, we obtain in the
nonrelativistic
limit
\begin{equation}
\label{nonr}
\frac{P^{2}}{2m}-\omega(k_{\vartheta}s
\pm\frac{1}{2})+\frac{m^2\omega^{2}%
\rho^{2}}{2}=\epsilon
\end{equation}
notice that Eq. (\ref{nonr}) corresponds to the Schr\"odinger
Hamiltonian of a
harmonic oscillator with an additional spin dependent term given by
$-\omega(k_{\vartheta}s \pm\frac{1}{2})$. This contribution is
proportional to
the frequency of the oscillator.

It would be interesting to analyze the Dirac oscillator in more complex
configurations where electromagnetic and gravitational interactions are
present, regretfully in this direction the possibilities of finding
exact
solvable examples are limited to those were the Dirac equation with
anomalous
moment are soluble. \cite{Bagrov}. A detailed discussion of this problem
will be
the objetive of a forthcoming publication


\begin{references}
\bibitem{Moshinsky} M. Moshinsky and A. Szczepaniak, J. Phys. A {\bf 22}
L817
(1989)
\bibitem{Benitez} J. Ben\'{\i}tez, R. P. Mart\'{\i}nez y Romero,
H. N. N\'u\~nez-Y\'epez, and A. L. Salas-Brito, Phys. Rev. Lett. {\bf
64}, 1643
(1990)
\bibitem{Dominguez} F. Dom\'{\i}nguez-Adame, Phys. Lett A. {\bf 162} 18
(1992)
\bibitem{Wilczek} F. Wilczek, Phys. Rev. Lett. {\bf 48} 1144 (1982)
\bibitem{Hagen} C. R. Hagen, Phys. Rev. Lett {\bf 64}, 503 (1990)

\bibitem{Shishkin} G. V. Shishkin and V. M. Villalba, J. Math. Phys.
{\bf 30}
2373 (1989)
\bibitem{Kamke}  E. Kamke, {\it Differentialgleichungen
L\"osungsmethoden und
L\"osungen} (Leipzig, 1959)
\bibitem{Abramowitz} {\it Handbook of Mathematical Functions,} Natl.
Bur.
Stand. Appl. Math. Ser. No. 55, edited by M. Abramowitz and
I. Stegun (U.S. GPO, Washington, D.C., 1965).
\bibitem{Bagrov}  V. G Bagrov, and D. Gitman, {\it Exact Solutions of
Relativistic Wave Equations} (Kluwer, Dordrecht, 1990)


\end{references}
\end{document}